\documentclass[sigconf]{acmart}

\usepackage{booktabs} 
\usepackage{multirow}
\usepackage{comment}
\usepackage{subfigure}

\usepackage{color}
\usepackage{soul}
\usepackage{hyperref}

 \hypersetup{
   colorlinks   = true, 
   urlcolor     = black, 
   linkcolor    = black, 
   citecolor   = black 
 }

\AtBeginDocument{%
  \providecommand\BibTeX{{%
    \normalfont B\kern-0.5em{\scshape i\kern-0.25em b}\kern-0.8em\TeX}}}

\begin{document}

\title{When HLS Meets FPGA HBM: Benchmarking \\ and Bandwidth Optimization}



\author{Young-kyu~Choi, Yuze~Chi, Jie~Wang, Licheng~Guo,     and~Jason~Cong}
\email{[ykchoi,chiyuze,jiewang,lcguo,cong]@cs.ucla.edu}
\affiliation{%
  \institution{Computer Science Department, University of California, Los Angeles, CA 90095}
}






\begin{abstract}
With the recent release of High Bandwidth Memory (HBM) based FPGA boards, developers can now exploit unprecedented external memory bandwidth.
This allows more memory-bounded applications to benefit from FPGA acceleration. However, we found that it is not easy to fully utilize the available bandwidth when developing some applications with high-level synthesis (HLS) tools. This is due to the limitation of existing HLS tools when accessing HBM board's large number of independent external memory channels. In this paper, we measure the performance of three recent representative HBM FPGA boards (Intel's Stratix 10 MX and Xilinx's Alveo U50/U280 boards) with microbenchmarks and analyze the HLS overhead. Next, we propose HLS-based optimization techniques to improve the effective bandwidth when a PE accesses multiple HBM channels or multiple PEs access an HBM channel. Our experiment demonstrates that the effective bandwidth improves by 2.4X-3.8X. We also provide a list of insights for future improvement of the HBM FPGA HLS design flow.
\end{abstract}






\maketitle

\section{Introduction}

Although field-programmable gate array (FPGA) is known to provide a high-performance and energy-efficient solution for many applications, there is one class of applications where FPGA is generally known to be less competitive: memory-bound applications (e.g., ~\cite{guo2019hardware,Chi2018,qiao2018high,fpga16-fpgp}). In a recent study \cite{Cong2018}, the authors report that GPUs typically outperform FPGAs in applications that require high external memory bandwidth. The Virtex-7 690T FPGA board used for the experiment reportedly has only 13~GB/s peak DRAM bandwidth, which is much smaller than the 290~GB/s bandwidth of the Tesla K40 GPU board used in the study (even though the two boards are based on the same 28~nm technology). This result is consistent with comparative studies for earlier generation of FPGAs and GPUs~\cite{Cooke2015,Cope2010}---FPGAs traditionally were at a disadvantage compared to GPUs for applications with low reuse rate. The FPGA DRAM bandwidth was also lower than the CPUs---Sandy Bridge E5-2670 (32~nm, similar generation as Virtex-7 in \cite{Cong2018}) has a peak bandwidth of 42~GB/s \cite{Molka2014}.

But with the recent emergence of High Bandwidth Memory 2 (HBM2)~\cite{Jedec:HBM} FPGA boards, it is possible that future FPGA will be able to compete with GPUs when it comes to memory-bound applications. Xilinx's Alveo U50~\cite{Xilinx:U50}, U280~\cite{Xilinx:U280}, and Intel's Stratix 10 MX \cite{Intel:HBM} have a theoretical bandwidth of about 400~GB/s (two HBM2 DRAMs), which approaches that of Nvidia's Titan V GPU~\cite{Nvidia:TitanV} (650~GB/s, three HBM2 DRAMs). With such high memory bandwidth, these HBM-based FPGA platforms have the potential to allow a wider range of applications to benefit from FPGA acceleration.

One of the defining characteristics of the HBM FPGA boards is the existence of independent and distributed HBM channels (e.g., Fig.~\ref{fig:s10arch}). To take full advantage of this architecture, programmers need to determine the most efficient way to connect multiple PEs to multiple HBM memory controllers. It is worth noting that Convey HC-1ex platform \cite{Bakos2010} also has multiple (64) DRAM channels like the HBM FPGA boards. The difference is that PEs in Convey HC-1ex issue individual FIFO requests of 64b data, but HBM PEs are connected to 256b/512b AXI bus interface. Thus, utilizing the bus burst access feature has a large impact on the performance of HBM boards. Also, Convey HC-1ex has a pre-synthesized full crossbar between PEs and DRAM, but FPGA programmers need to customize the interconnect in the HBM boards.

\begin{table}[h]
\caption{Effective bandwidth of memory-bound applications on Stratix 10 MX and Alveo U280 using HLS tools}
\label{tab:app_perf}
\centering
\begin{tabular}{|ccc|ccc|}
\hline
\multirow{2}{*}{App}&\multirow{2}{*}{Plat}&PC&KClk&EffBW&EffBW/PC\\
&&\#& (MHz)&(GB/s)&(GB/s)\\
\hline
\multirow{3}{*}{MV}&S10&32&418&372&11.6\\
&U50&24&300&317&13.2\\
&U280&28&274&370&13.2\\
\hline
\multirow{3}{*}{Stencil}&S10&32&260&246&7.7\\
&U50&16&300&203&12.7\\
&U280&16&293&206&12.9\\
\hline
\multirow{2}{*}{Bucket}&S10&16&287&137&8.6\\
\multirow{2}{*}{sort}&U50&16&300&36&2.3\\
&U280&16&300&36&2.3\\
\hline
\multirow{2}{*}{Binary}&S10&32&310&5.2&0.16\\
\multirow{2}{*}{search}&U50&24&300&6.6&0.27\\
&U280&28&300&7.7&0.28\\
\hline
\end{tabular}
\end{table}

Table~\ref{tab:app_perf} shows the effective bandwidth of memory-bound applications\footnote{We only tested designs utilizing power-of-2 HBM PCs for stencil and bucket sort. The reason for using less than 32 PCs in Alveo U50/U280 will be explained in Section~\ref{sec:long_seq}. We only used 16 PCs in Stratix 10 MX for bucket sort due to high resource consumption and routing complexity (more details in Section~\ref{sec:many2many}).} we have implemented on the HBM boards. The kernels are written in C (Xilinx Vivado HLS \cite{Xilinx:VivadoHLS}) and OpenCL (Intel AOCL \cite{Intel:OpenCLProgramming}) for ease of programming and faster development cycle \cite{Lahti2019}. For dense matrix-vector multiplication (MV) and stencil, the effective bandwidth per pseudo channel (PC) is similar\footnote{The effective bandwidth of stencil and bucket sort on Stratix 10 MX has been slightly reduced (7.7\textasciitilde8.6~GB/s) due to the low kernel frequency (260\textasciitilde287~MHz). This will be further explained in Section~\ref{sec:freq_variation}.} to the boards' sequential access bandwidth (Section~\ref{sec:long_seq}). Both applications can evenly distribute the workload among the available HBM PCs, and their long sequential memory access pattern allows a single processing element (PE) to fully saturate an HBM PC's available bandwidth.

However, the effective bandwidth is far lower for bucket sort and binary search. In bucket sort, a PE distributes keys to multiple HBM PCs (one HBM PC corresponds to one bucket). Alveo U280 provides an area-efficient crossbar to facilitate this multi-PC distribution. But, as will be explained in Section~\ref{sec:opt_description}, it is difficult to enable external memory burst access to multiple PCs in the current high-level synthesis (HLS) programming environment. For binary search, its latency-bound characteristic is the main reason for the reduced bandwidth in both platforms. But whereas Stratix 10 MX allows multiple PEs to share access to an HBM PC to hide the latency, it is difficult to adopt a similar architecture in Alveo U280 due to HLS limitation (more details in Section~\ref{sec:multiPEtoPC}).

In this paper, we will first evaluate the performance of three recent representative HBM2 FPGA boards--Intel Stratix 10 MX~\cite{Intel:HBM} and Xilinx Alveo U50~\cite{Xilinx:U50} and U280~\cite{Xilinx:U280}--with various memory access patterns in Section~\ref{sec:evaluation}. Based on this evaluation, we identify several problems in using existing commercial HLS tools for HBM application development.
A novel HLS-based approach will be presented in Section~\ref{sec:opt} to improve the effective bandwidth when a PE accesses several PCs or when several PEs share access to a PC. The opportunities for future research will be presented in Section~\ref{sec:insight}.

A related paper named Shuhai~\cite{Wang2020} has been recently published in FCCM'20. Shuhai is a benchmarking tool used to evaluate Alveo U280's HBM and DRAM. It measures the memory throughput and latency for various burst size and strides using RTL microbenchmarks. Whereas Shuhai predicted the per-PC bandwidth to linearly scale up to all PCs, we show that such scaling cannot be achieved in practice and provide an explanation. We also quantify the overhead of using HLS-based design flow compared to Shuhai's RTL-based flow. Moreover, we not only evaluate Alveo U280 but also compare the performance and architectural differences with Stratix 10 MX board.

We make the following contributions in this paper:

\begin{itemize}
    \item
We quantify the performance of several memory-bound applications on the latest Intel and Xilinx FPGA HBM boards and identify problems in directly applying existing commercial HLS tools to develop memory-bound applications.
    \item
With microbenchmarks, we analyze the cause for the performance degradation when using HLS tools.
	\item
We propose a novel HLS-based solution for Alveo U280 to increase the effective bandwidth when a PE accesses several PCs or when several PEs share access to a PC.
    \item
We present several insights for the future improvement of the FPGA HBM HLS design flow.
\end{itemize}

The benchmarks used in this work can be found in: \\ \href{https://github.com/UCLA-VAST/hbmbench}{https://github.com/UCLA-VAST/hbmbench}.

\section{Background}
\label{sec:HBM}

\subsection{High Bandwidth Memory 2}

High Bandwidth Memory \cite{Jedec:HBM} is a 3D-stacked DRAM designed to provide a high memory bandwidth. Each stack is composed of 2\textasciitilde8 HBM dies and 1024 data I/Os. The HBM dies are connected to a base logic die using Through Silicon Via (TSV) technology. The base logic die connects to FPGA/GPU/CPU dies through an interposer. The maximum I/O data rate improves from 1~Gbps in HBM1 to 2~Gbps in HBM2. This is partially enabled by the use of two pseudo channels (PCs) per physical channel to hide the latency \cite{Intel:HBM,Jun2017}. Sixteen PCs exist per stack, and they can be accessed independently.

\subsection{FPGA Platforms for HBM2}
\label{sec:fpga_platforms}

\subsubsection{Intel Stratix 10 MX}
\label{sec:S10MX}

\begin{figure}[t]
\centering
\includegraphics[width=0.99\linewidth]{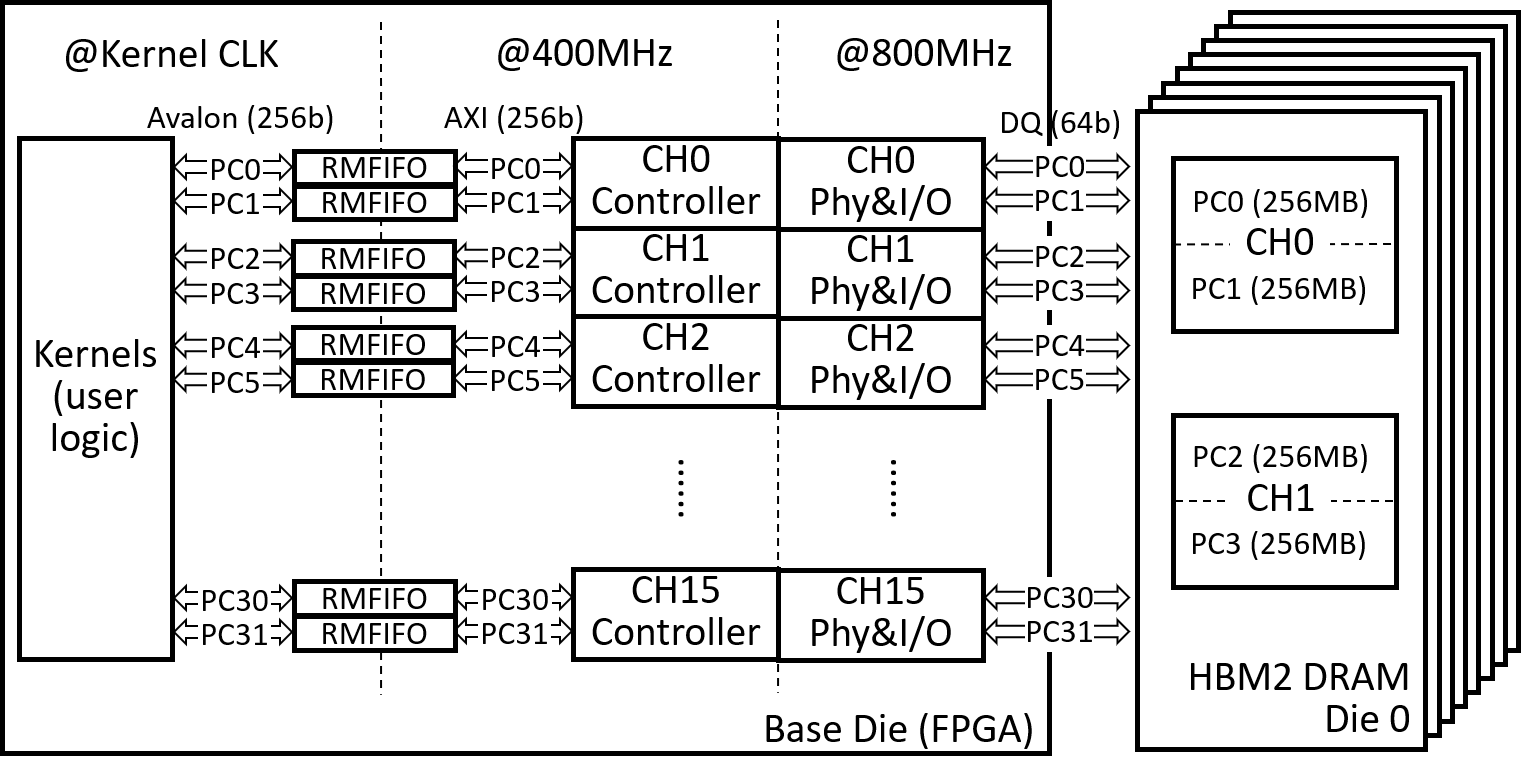}
\caption{Intel Stratix 10 MX Architecture \cite{Intel:HBM}}
\label{fig:s10arch}
\end{figure}

The overall architecture of Intel Stratix 10 MX is shown in Fig.~\ref{fig:s10arch} \cite{Intel:HBM}. Intel Stratix 10 MX (early-silicon version) consists of an FPGA and two HBM2 stacks (8 HBM2 dies). The FPGA resource is presented in Table~\ref{tab:constraint}. The FPGA and the HBM2 dies are connected through 32 independent pseudo channels (PCs). Each PC has 256MB of capacity (8GB in total). Each PC is connected to the FPGA PHY layer through 64 data I/Os that operates at 800MHz (double data rate). The data communication between the kernels (user logic) and the HBM2 memory is managed by the HBM controller (HBMC). AXI4 \cite{ARM:AXI} and Avalon \cite{Intel:Avalon} interfaces with 256 data bitwidth are used to communicate with the kernel side.

\begin{table}[ht]
\caption{FPGA resource on Stratix 10 MX, Alveo U50, and Alveo U280}
\label{tab:constraint}
\centering
\begin{tabular}{|c|c c c c|}
\hline
Platform&LUT & FF & DSP & BRAM \\
\hline
S10 MX & 1.41M & 2.81M & 6.85K & 3.95K \\
Alv U50 & 872K & 1.74M & 5.96K & 2.69K \\
Alv U280 & 1.30M & 2.60M & 9.02K & 4.03K \\
\hline
\end{tabular}
\end{table}

The clock frequency of kernels may vary (capped at 450MHz) depending on the complexity of the user logic. Since the frequency of HBMCs is fixed to 400MHz, rate matching (RM) FIFOs are inserted between the kernels and the memory controllers. The ideal memory bandwidth is 410GB/s (=~256b * 32PCs * 400MHz =~64b * 32PCs * 2 * 800MHz).

\subsubsection{Xilinx Alveo U50 and U280}
\label{sec:AlvU50}

\begin{figure}[t]
\centering
\includegraphics[width=0.9\linewidth]{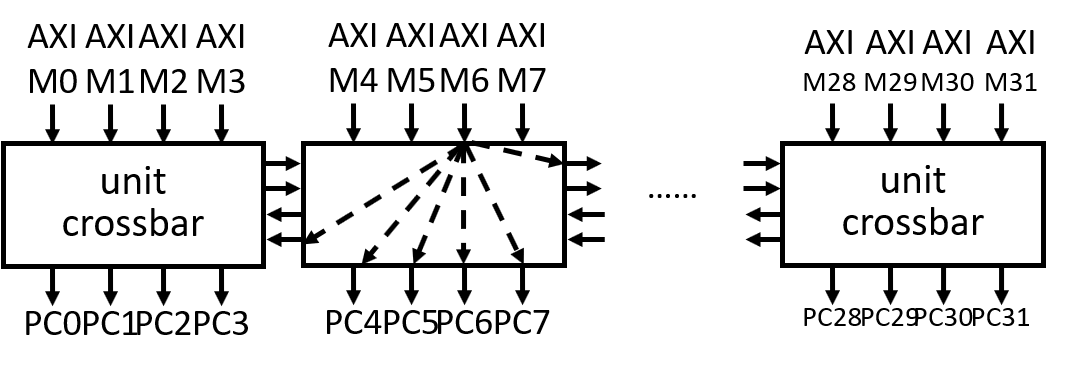}
\caption{AXI crossbar in Alveo U50 and U280 \cite{Xilinx:HBMIP}}
\label{fig:axi_crossbar}
\end{figure}

Similar to Stratix 10 MX, Xilinx Alveo U50 FPGA has an FPGA and two HBM2 stacks~\cite{Xilinx:U50}. There are 32 PCs, each with 64b data I/Os and 256MB capacity. The FPGA is composed of two super logic regions (SLRs), and its resource is shown in Table~\ref{tab:constraint}.
The two HBM2 stacks are physically connected to the bottom SLR (SLR0).
Data I/Os runs at 900MHz (double data rate), and the the total ideal memory bandwidth is 460GB/s (=~64b*32PCs*2*900MHz).
An HBMC IP slave interfaces the kernel (user logic) via a 256b AXI3 interface running at 450MHz~\cite{Xilinx:HBMIP}. A kernel master has a 512b AXI3 interface that may run up to 300~MHz. Four of AXI masters and four of AXI slaves are connected through a fully-connected crossbar (Fig.~\ref{fig:axi_crossbar}). There exists a datapath across the 4$\times$4 unit crossbar, but the bandwidth may be limited due to the network contention among the switches~\cite{Xilinx:HBMIP}. The thermal design power (TDP) of U50 is 75W, and it may not be possible to utilize all HBM channels and FPGA logic due to the power restriction~\cite{Xilinx:U50}.

Alveo U280 has a similar architecture as U50 except that its FPGA is composed of three SLRs~\cite{Xilinx:U280}. Its TDP is 200W. Note that Alveo U280 also has traditional DDR DRAM---but we decided not to utilize the traditional DRAM because the purpose of this paper is to evaluate the HBM memory. Readers are referred to \cite{Miao2019} for optimization case studies on heterogeneous external memory architectures.

\subsection{HLS Programming for HBM2}
\label{sec:hbmhls}

For Stratix 10 MX, we program kernels in OpenCL and synthesize them using Intel's Quartus \cite{Intel:Quartus} and AOCL \cite{Intel:OpenCLProgramming} 19.4 tools. For Alveo U50 and U280, we program in C and utilize Xilinx's Vitis~\cite{Xilinx:Vitis} and Vivado HLS~\cite{Xilinx:VivadoHLS} 2019.2 tools. We use dataflow programming style (C functions executing in parallel and communicating through streaming FIFOs) for Alveo kernels to achieve high throughput with small BRAM consumption \cite{Xilinx:VivadoHLS}.

\subsubsection{Accessing Multiple PCs from a PE}
\label{sec:multiPCtoPE}

\begin{figure}[t]
\centering
\includegraphics[width=0.99\linewidth]{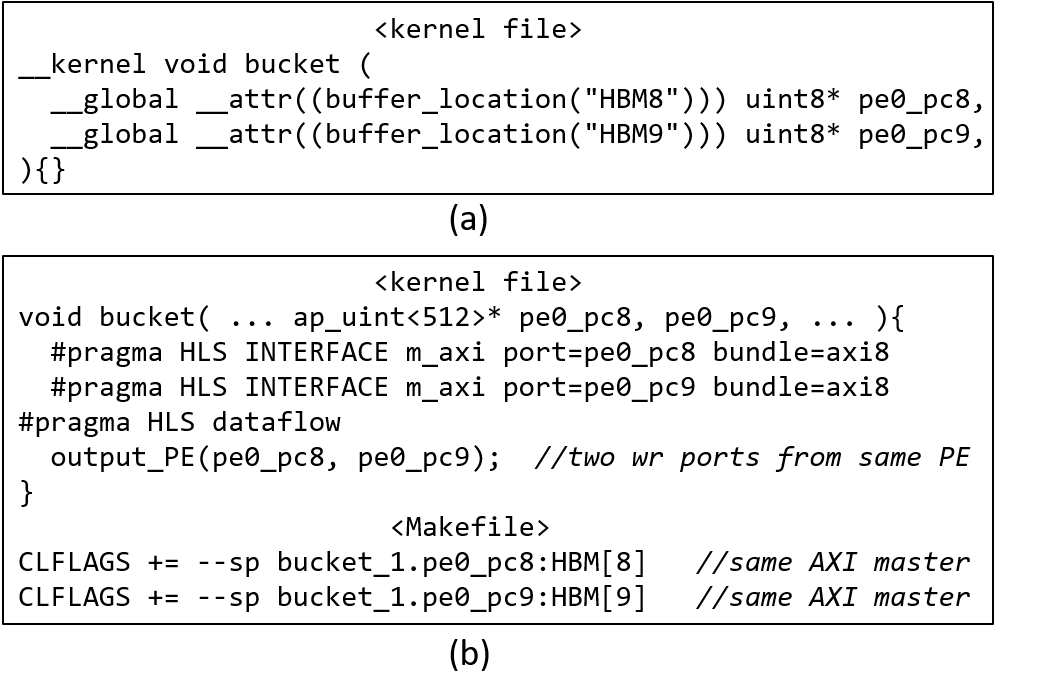}
\caption{Assigning different destination PCs to a bucket sort top function argument (a) AOCL style for Stratix 10 MX (b) Vivado HLS style for Alveo U50/U280}
\label{fig:hls_bucket}
\end{figure}

On Stratix 10 MX, programmers specify the target PC for each function argument (PE's port) using ``buffer\_location" attribute (Fig.~\ref{fig:hls_bucket}(a)). This creates a new connection between a PE and an AXI master of a PC. Although programmer-friendly, the inter-connection may consume most of the FPGA resource (Section~\ref{sec:many2many}).

On Alveo U50 and U280, programmers can exploit the built-in AXI crossbar to avoid high resource consumption. An example HLS code is provided in Fig.~\ref{fig:hls_bucket}(b). Given a single function (PE) with multiple arguments (ports), a programmer can assign different PC number (PC8 and PC9) to each argument while assigning the same bundle (axi8) to all arguments. A \textit{bundle} is a Vivado HLS concept that corresponds to an AXI master. When an AXI master is connected to ports with multiple PCs (see Makefile of Fig.~\ref{fig:hls_bucket}(b)), it will automatically use the AXI crossbar. Although area-efficient, directly using such assignment strategy for bucket sort (Table~\ref{tab:app_perf}) may result in severe performance degradation (details in Section~\ref{sec:opt_description}).

\subsubsection{Accessing a PC from Multiple PEs}
\label{sec:multiPEtoPC}

On Stratix 10 MX, ports from multiple PEs may have the same buffer\_location attribute and access the same PC. Even if a PE port's access rate is low (e.g., binary search), one could increase the bandwidth utilization by connecting several PEs to the same PC.

\begin{figure}[t]
\centering
\includegraphics[width=0.99\linewidth]{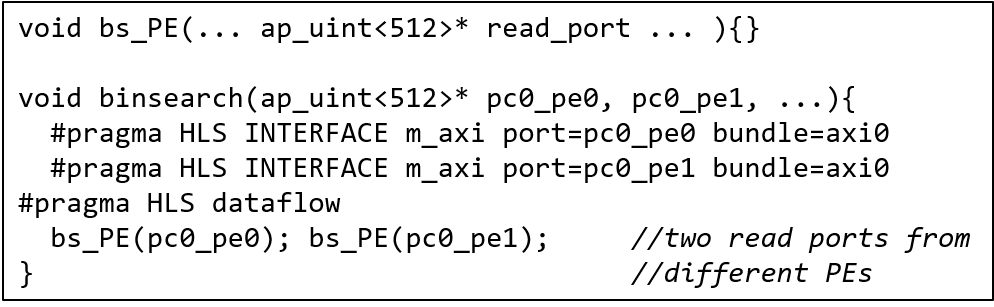}
\caption{Illegal Vivado HLS coding style of two read ports from different PEs connected to the same bundle}
\label{fig:hls_bsearch}
\end{figure}

Alveo U50 and U280, on the other hand, allow only one read PE and one write PE to be connected to a bundle in dataflow.\footnote{The restriction is on the number of connected PEs (functions), not the number of ports (function arguments) from a PE. Multiple read/write ports from a single PE may be connected to a bundle (e.g. to access multiple PCs as in Section~\ref{sec:multiPCtoPE}).} An example of illegal coding style due to two read PEs connected the same bundle is shown in Fig.~\ref{fig:hls_bsearch}. This limitation was not a problem in non-HBM boards (e.g., \cite{Alphadata:KU3,Amazon:F1}) because one could easily utilize multiple bundles for access to a few (1~\textasciitilde~4) DRAM channels. Assigning multiple bundles per DRAM channel allows parallel DRAM access from multiple PEs in non-HBM boards. But due to the fixed number of AXI masters in Alveo U50 and U280, the number of usable bundles is fixed to 30 for U50 and 32 for U280. This effectively limits one bundle to be used for one PC. Thus, a PC's bandwidth cannot be fully utilized unless one read PE and one write PE can make full use of it.

\subsection{Applications}
\label{sec:apps}

\subsubsection{Bucket Sort}
\label{sec:bsort}

We sort an array of keys that would be sent to buckets. Each bucket is stored in a single HBM PC, and this allows a second stage of sorting (e.g., with merge sort) to be independently performed for each bucket. For simplification, we assume a key is 512b/256b (Alveo U280/Stratix 10 MX) long, and there is an equal number of keys that would be sent to each bucket. We fetch keys from 8 read HBM PCs and send them to the buckets among 8 write HBM PCs. We use the many-to-many unicast architecture that will be described in Section~\ref{sec:many2many}.

\subsubsection{Radix Sort}
\label{sec:rsort}

Radix sort is composed of multiple iterations---each iteration sorting based on 3 (=log8) bits of the key. We sort from least significant bit to most significant bit to ensure stability of the sort. Similar to bucket sort, we use 8 read PCs and 8 write PCs for radix sort. We switch the input and output PCs in each iteration and send the 512b/256b keys in a ping-pong fashion.

\subsubsection{Binary Search}
\label{sec:bsearch}

We perform a binary search on an array with the size of 16MB. Each data element is set to 512b/256b. Each PE accesses one PC, and multiple PEs executes the search independent of each other.

\subsubsection{Depth-first Search}
\label{sec:dfs}

We conduct depth-first search on a binary tree implemented as a linked list. The value of each node and ID of the connected nodes form 512b/256b data. Each PE has a stack to store the address of the nodes to be searched later.

\begin{table}[t]
\caption{Summary of HBM2 FPGA platform evaluation result}
\label{tab:summary}
\centering
\begin{tabular}{|c c|c c c|}
\hline
\multicolumn{2}{|c|}{Platform}&S10 MX&Alv U280&Alv U280 \\
\hline
\multicolumn{2}{|c|}{Ker Lang}&OpenCL&C&RTL \cite{Wang2020}\\
\hline
\multicolumn{2}{|c|}{Seq R/W BW (GB/s)} & 357 & 388 & 425 \\
\hline
Strided & 0.25KB & 182 & 79 & 420\\
R/W BW & 1KB & 158 & 79 & 420\\
(GB/s) & 4KB & 50 & 42 & 70\\
\hline
Read seq & 128b & 215 & 141 & \\
BW top & 256b & 353 & 276 & N/A \\
arg width & 512b & 369 & 391 & \\
\hline
Read seq & 150MHz & 154 & 285 & \\
BW kernel & 200MHz & 205 & 380 & N/A \\
clk freq & 250MHz & 256 & 393 & \\
\hline
\multicolumn{2}{|c|}{Read latency (ns)} & 633 & 229 & 107 \\
 \hline
M2M unicast&2$\times$2& 186 & 209 & \multirow{3}{*}{N/A} \\
16 CH BW & 4$\times$4 & 185 & 209 &  \\
(GB/s) & 8$\times$8 & 175 & 96 &  \\
\hline
\end{tabular}
\end{table}

\section{HBM2 FPGA Platform Evaluation}
\label{sec:evaluation}

In this section, we will analyze the on-board behavior of the HBM2 memory system using a set of HLS microbenchmarks. The summary of the result is shown in Table~\ref{tab:summary}. 
When applicable, we also make a quantitative comparison with an RTL-based evaluation result in Shuhai \cite{Wang2020}. Note that, except for Section~\ref{sec:long_seq}, we omit the result for Alveo U50 due to space limitation---per-PC performance of Alveo U50 is similar to that of U280.

\subsection{Sequential Access Bandwidth}
\label{sec:long_seq}

\begin{figure}[t]
\centering
\includegraphics[width=0.99\linewidth]{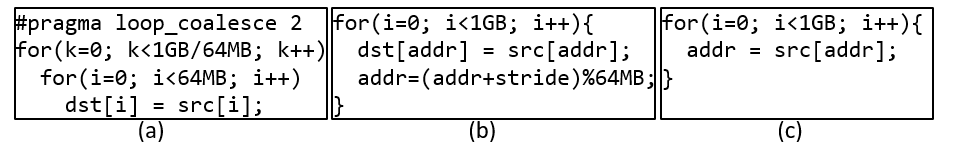}
\caption{Microbenchmark code for (a) sequential access bandwidth (b) strided access bandwidth (c) read latency measurement}
\label{fig:microbench}
\end{figure}

The maximum memory bandwidth of the HBM boards is measured with a sequential access pattern shown in Fig.~\ref{fig:microbench}(a). The experiment performs a simple data copy with read \& write, read only, and write only operations. We use the default bus data bitwidth of 256b for Stratix 10 MX and 512b for Alveo U50 and U280. 
In order to reduce the effect of kernel invocation overhead, we transfer 1~GB of data per PC. Since a single PC cannot store 1~GB of contiguous data, we repeatedly ($k$ loop) copy 64MB of data ($i$ loop).
We coalesce (flatten) the $k$ and the $i$ loops to increase the burst length and to remove the loop iteration latency of the $i$ loop.

Shuhai \cite{Wang2020} assumes that the total effective bandwidth can be estimated by multiplying the bandwidth of a single PC by the number of PCs. In practice, however, we found that it is difficult to utilize all PCs. PC 30 and 31 cannot be used since they overlap with the PCIE static region~\cite{Xilinx:U50}. Also, Vitis was not able to complete the routing for PC 24-29 for U50 due to the congestion near the HBM ports. Thus, we could only use 24~PCs in U50 and 30~PCs in U280 for this experiment. 
Adding more computation logic slightly worsens this problem and we can typically use 28~PCs for Alveo U280 (MV and binary search in Table~\ref{tab:app_perf}).

\begin{table}[ht]
\caption{Effective memory bandwidth with sequential access pattern (GB/s)}
\label{tab:bw_seq}
\small
\centering
\begin{tabular}{|c|c|c c c|c|}
\hline
Platform & PC\# & Read \& Write & Read only & Write only & Ideal \\
\hline
\multirow{2}{*}{S10 MX} & 32 & 357 & 353 & 354 & 410\\
& 1 & 11.2 & 11.0 & 11.1 & 12.8\\
\hline
\multirow{2}{*}{Alv U50} & 24 & 310 & 316 & 314 & 460\\
& 1 & 12.9 & 13.2 & 13.1 & 14.4\\
\hline
Alv U280 & 30 & 388 & 391 & 393 & 460\\
(This work) & 1 & 12.9 & 13.0 & 13.1 & 14.4\\
\hline
Alv U280 & 32 & 425 & N/A & N/A & 460\\
\cite{Wang2020} & 1 & 13.3 & N/A & N/A & 14.4\\
\hline
\end{tabular}
\end{table}

The overall effective bandwidth and the effective bandwidth per PC are presented in Table~\ref{tab:bw_seq}. The per-PC bandwidth is 15\% higher in U50 and U280 compared to Stratix 10 MX, because the Alveo boards use faster HBM PHY clock of 900MHz (800MHz in Stratix 10 MX). Also, the per-PC bandwidth result shows that we can obtain about 87\% of the ideal bandwidth for Stratix 10 MX and 90\% for Alveo U50/U280. The bandwidth can be saturated with read-only or write-only access in all boards.

Due to the limitation in the internal memory or the application's memory access pattern, the burst length may be far shorter than 64MB. Readers may refer to \cite{Choi2016,Choi2017+,Choi2019,Park2004} on how the effective bandwidth changes depending on the burst length and the memory latency.

\subsection{Strided Access Bandwidth}
\label{sec:stridebw}

We measure the effective bandwidth when accessing data with a fixed address stride. The granularity of the data is set to 512b.
The microbenchmark is shown in Fig.~\ref{fig:microbench}(b).

\begin{figure}[t]
\centering
\subfigure[]{\includegraphics[width=0.48\linewidth]{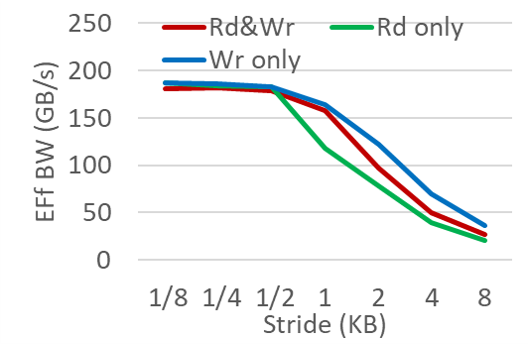}}
\subfigure[]{\includegraphics[width=0.48\linewidth]{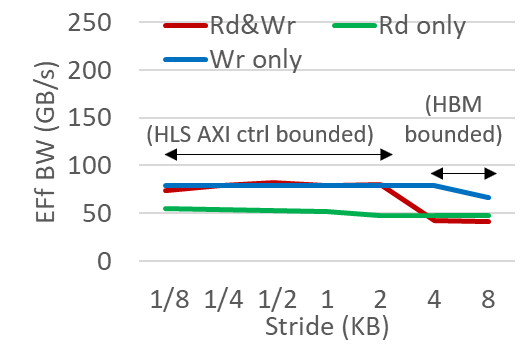}}
\caption{Effective memory bandwidth with varying stride (a) Stratix 10 MX (b) Alveo U280}
\label{fig:stride_bw}
\end{figure}

The result for Alveo U280 is shown in Fig.~\ref{fig:stride_bw}(b). Compared to the strided access bandwidth reported in Shuhai \cite{Wang2020} (about 420~GB/s for both 0.25KB and 1KB stride, Table~\ref{tab:summary}), the obtained effective bandwidth is about 5X lower (79~GB/s for both 0.25KB and 1KB stride). The reason can be found in how the memory request is sent to the AXI bus. In Shuhai, a new memory request is sent as soon as the AXI bus is ready to receive a new request. In HLS, the generated AXI master has some limitation in bookkeeping the outstanding requests. Since a burst access cannot be used for strided access, the number of outstanding requests becomes much larger than that of sequential access. As a result, HLS AXI master often stalls and the performance is limited to about 50--80~GB/s. When the stride is longer than 4KB, the effective bandwidth is limited by the latency of opening a new page in the HBM memory \cite{Wang2020}, and the bandwidth becomes similar to that of Shuhai. 

The effective bandwidth of Stratix 10 MX, on the other hand, degrades more gracefully with larger stride (Fig.~\ref{fig:stride_bw}(a)). Stratix 10 MX fixes the AXI burst length to one \cite{Intel:HBM} and instead relies on multiple outstanding requests even for sequential access. Thus, we can deduce that Stratix 10 MX's HLS AXI master is designed to handle more outstanding requests, and it can better retain the effective bandwidth even with short accesses.

\begin{figure}[t]
\centering
\subfigure[]{\includegraphics[width=0.48\linewidth]{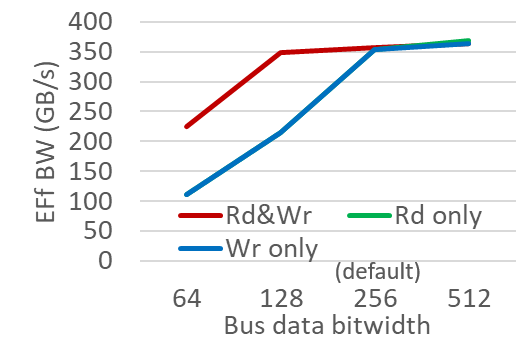}}
\subfigure[]{\includegraphics[width=0.48\linewidth]{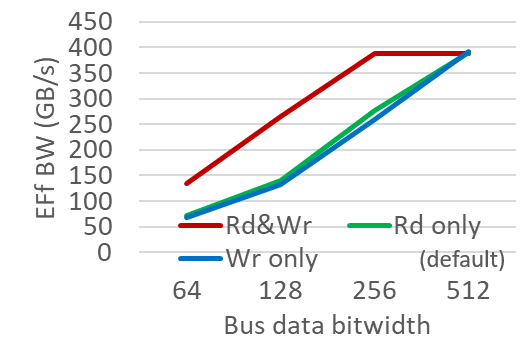}}
\caption{Effective memory bandwidth with varying kernel top function argument bitwidth (a) Stratix 10 MX (b) Alveo U280}
\label{fig:bitwidth_result}
\end{figure}

\subsection{Sequential Access Bandwidth with Varying Kernel Top Function Argument Bitwidth}

We test the effective bandwidth of sequential access after varying the bitwidth of the kernel's top function argument. This changes the data width of the AXI bus being utilized. Fig.~\ref{fig:bitwidth_result} (a) shows that Stratix 10 MX bandwidth can be saturated with 128b variable when simultaneously reading and writing. It can also be saturated with 256b bus data width with read or write only. In Alveo U280, however, the performance is saturated with wider 512b data width (read or write only). The reason for the difference is that the Stratix 10 MX kernel can be typically synthesized at a higher frequency (the maximum is 450MHz for S10 MX and 300MHz for Alv U50/U280). As a result, the Alveo U50/U280 kernels require larger bus data width to saturate the bandwidth. This suggests that Alveo U50/U280 HLS programmers would need to use larger 512b data types for top function argument and distribute it to more PEs after breaking it down to primitive data types (e.g., short or int). The complexity of the data distribution may reduce the kernel frequency.

\subsection{Sequential Access Bandwidth with Kernel Frequency Variation}
\label{sec:freq_variation}

The HBM2 controller and the PHY layer transfer data at a fixed frequency, but the kernel clock may change depending on the complexity of the user logic. If the kernel clock is too slow, the kernel may not be able to saturate the maximum bandwidth. To obtain the bandwidth-saturating kernel clock frequency, we measure the effective bandwidth of the sequential access pattern with varying kernel clock frequency. We fix the bus data width to default 256b for Stratix 10 MX and 512b for Alveo U280. Although Vitis allows the kernel clock frequency to be freely configured after bitstream generation, Quartus does not have such functionality. Thus, for Stratix 10 MX, we present the estimated bandwidth assuming ideal bandwidth, saturated by the sequential bandwidth (from Table~\ref{tab:bw_seq}).

\begin{figure}[t]
\centering
\subfigure[]{\includegraphics[width=0.48\linewidth]{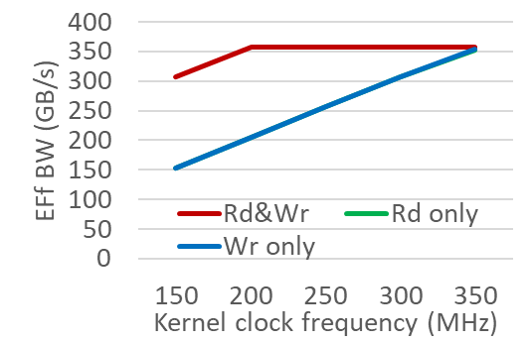}}
\subfigure[]{\includegraphics[width=0.48\linewidth]{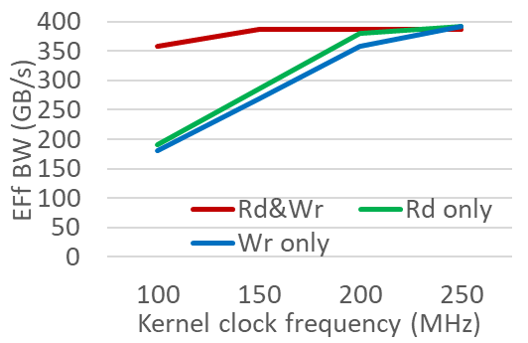}}
\caption{Effective bandwidth with different kernel frequency (a) Stratix 10 MX (estimated) (b) Alveo U280 (experimented)}
\label{fig:freq_result}
\end{figure}

Fig.~\ref{fig:freq_result}(a) shows that the maximum bandwidth is reached at about 200MHz for read \& write and 350MHz for read only and write only on Stratix 10 MX. This is consistent with the Stratix 10 MX performance result of stencil and bucket sort in Table~\ref{tab:app_perf}---these applications use read or write only ports and have kernel frequency of 260\textasciitilde287~MHz. From Fig.~\ref{fig:freq_result}(a), it is not possible to saturate the bandwidth at this frequency. 

The Alveo U280 experimental result in Fig.~\ref{fig:freq_result}(b) shows that the bandwidth is saturated at about 150MHz for read \& write and 200MHz for read only and write only. The saturation point is reached at a lower frequency on Alveo U50/U280, because it uses a larger top function argument bitwidth. The different saturation frequency may make Intel Stratix 10 MX to suffer more from complex circuitry. But also note that the maximum kernel frequency is 450~MHz on Stratix 10 MX and 300~MHz on Alveo U50/U280. 

\subsection{Many-to-Many Unicast Bandwidth}
\label{sec:many2many}

Processing elements may communicate with HBM2 PCs in various patterns---including scatter, gather, and broadcast. Among them, we analyze the most complex communication pattern where all PEs communicate with all PCs simultaneously.

We quantify the FPGA resource consumption and the effective bandwidth for transferring data from 8 read PCs to 8 write PCs. Data in a read PC is written to 1\textasciitilde8 different write PCs. There are 8 PEs transferring the data in parallel. Each PE transfers data from a single read PC to a single write PC at a time (many-to-many unicast). If read data needs to be written to multiple PCs, the write PC is changed in round-robin. The complexity of the communication architecture increases from 1$\times$1 (data read from one PC is written to one PC) to 8$\times$8 (data read from one PC is written to 8 PCs). Equal amount of data is read and written to each PC in a contiguous chunk.

\begin{table}[ht]
\caption{Effective bandwidth and resource consumption of many-to-many unicast (from 8 read PCs to 8 write PCs)}
\label{tab:crossbar}
\centering
\begin{tabular}{|c c|c|c c|}
\hline
\multirow{2}{*}{Plat}&Comm&FPGA Resource&KClk&EffBW\\
&compl&LUT / FF / DSP / BRAM & (MHz)&(GB/s)\\
\hline
& 1$\times$1 & 147K / 376K / 12 / 958 & 406 & 187\\
S10& 2$\times$2 & 157K / 422K / 12 / 1.1K & 438 & 186\\
MX& 4$\times$4 & 169K / 464K / 12 / 1.3K & 425 & 185\\
& 8$\times$8 & 193K / 559K / 12 / 1.6K & 385 & 175\\
\hline
& 1$\times$1 & 42K / 106K / 0 / 124 & 300 & 209\\
Alv& 2$\times$2 & 44K / 109K / 0 / 124 & 300 & 209\\
U280& 4$\times$4 & 47K / 112K / 0 / 124 & 300 & 209\\
& 8$\times$8 & 51K / 120K / 0 / 124 & 300 & 96\\
\hline
\end{tabular}
\end{table}

Experimental result is shown in Table.~\ref{tab:crossbar}. As explained in Section~\ref{sec:multiPCtoPE}, Stratix 10 MX creates a new connection from a PE to AXI masters proportional to the number of PCs accessed from each PE. A corresponding control logic is added as well. This causes the FPGA resource consumption to rapidly increase with a more complex communication scheme. As a result, we were not able to test the bandwidth with a 16$\times$16 crossbar that accesses all 32 PCs (16 read and 16 write PCs)---routing failed due to the complexity of the custom crossbar and the large resource consumption.

The difference in resource consumption for Alveo U280 is relatively small (only 9K LUTs and 14K FFs) even though the complexity has increased from 1$\times$1 to 8$\times$8. This is attributed to the built-in crossbar that connects all user logic AXI masters to all HBM PC AXI slaves (Section~\ref{sec:AlvU50}). Table.~\ref{tab:crossbar} also reveals that the effective bandwidth is maintained in 2$\times$2 and 4$\times$4 and drops rapidly in 8$\times$8. This is due to the data congestion when crossing the 4$\times$4 unit switch in the AXI crossbar.

\subsection{Memory Latency}
\label{sec:latency}

\textit{Memory latency} is defined as a round-trip delay between the time user logic makes a memory request and the time acknowledgement is sent back to the user logic. This becomes an important metric for latency-sensitive applications. We measure the read memory latency with a pointer-chasing microbenchmark shown in Fig.~\ref{fig:microbench}(c). The data is used as an address of the next element.

The measurement result is presented in Table~\ref{tab:mem_lat}. We break down the averaged total latency into the latency caused by the HLS PE and the latency of the memory system (HBMC + HBM memory). The HLS PE latency depends on the application. In Alveo U280, the HLS PE latency was obtained by observing the waveform. We were not able to observe the waveform in current Quartus debugging environment for Stratix 10 MX, so we present an estimate based on the loop latency in the AOCL synthesis report.

The overall latency is longer in Stratix 10 MX compared to Alveo U280 because of the heavy pipeline AOCL automatically applies to improve the throughput. As a side effect, the the latency of the pointer chasing PE becomes very long (492~ns). Such long latency also causes the the binary search application (Table~\ref{tab:app_perf}) to have a low effective bandwidth (5.2~GB/s). The HLS PE latency may be improved if HLS vendors provide optional latency-sensitive (less pipelined) flow.

\begin{table}[ht]
\caption{Read memory latency measurement result}
\label{tab:mem_lat}
\centering
\begin{tabular}{|c|c c|}
\hline
& S10 MX & Alv U280 \\
\hline
Total & 633~ns & 229~ns \\
\hline
HLS PE & ~\textasciitilde492~ns & 47~ns \\
HBMC+HBM & ~\textasciitilde141~ns & 182~ns \\
\hline
\hline
HBMC+HBM \cite{Wang2020} & N/A & 107~ns \\
\hline
\end{tabular}
\end{table}

Note that, compared to the read latency measurement in Shuahai~\cite{Wang2020} (107~ns), the latency of the HBMC and HBM on Alveo U280 is longer in our measurement (182~ns). This is likely due to the fact that Shuhai obtained page hit and disabled AXI crossbar~\cite{Wang2020}---which removes the crossbar latency and any interaction with other AXI masters accessing HBM in parallel.

\section{Effective Bandwidth Improvement for HBM HLS}
\label{sec:opt}

\subsection{Problem Description}
\label{sec:opt_description}

We identify two problems from the HLS-based implementation of memory-bound applications listed in Table~\ref{tab:app_perf}. First, in bucket sort, we use the many-to-many unicast architecture (Section~\ref{sec:many2many}) to distribute the keys from input PCs to output PCs (each PC corresponds to a bucket) in parallel. However, there is a large difference in effective bandwidth in bucket sort (36~GB/s, Table~\ref{tab:app_perf}) and many-to-many unicast microbenchmark result (96~GB/s, Table~\ref{tab:crossbar}). This is because in many-to-many unicast experiment, we transferred data from input PCs to output PCs in a large contiguous chunk. But, in bucket sort, there is no guarantee that two consecutive keys will be sent to the same bucket. This could become problematic since existing HLS tools do not automatically hold the data in buffer for burst AXI access to each HBM PC. Stratix 10 MX better retains a high bandwidth for short memory access (Section~\ref{sec:stridebw}), so the effective bandwidth is high (125~GB/s, Table~\ref{tab:app_perf}) for bucket sort. But short memory access does become a problem with Alveo U280 due to its dependence on long burst length---we found that Vivado HLS conservatively sets the AXI burst length to one for bucket sort key write.

Second, in binary search, some degradation in effective bandwidth is unavoidable because it needs to wait for data to arrive before it can access the next address. We can alleviate this problem by connecting several PEs to the same PC for shared access and amortize the memory latency. This technique increases the effective bandwidth from 3.1~GB/s to 5.2~GB/s for Stratix 10 MX. But, as mentioned in Section~\ref{sec:multiPEtoPC}, the HLS tool for Alveo U280 only allows up to one read and one write PE to access the same PC. As a result, the effective bandwidth could not be improved beyond 7.7~GB/s (Table~\ref{tab:app_perf}) for Alveo U280.

The two problems on Alveo U280 can be summarized as follows:

\begin{itemize}
    \item
Problem 1: Suppose there is a PE that accesses multiple PCs. The order of the destination PC is random but the access address to each PC is sequential. Then how can we improve the effective bandwidth in HLS?
    \item
Problem 2: Given multiple PEs with low memory access rate and an AXI master that allows read access from one PE and write access from one PE, how can we improve the effective bandwidth in HLS?
\end{itemize}

In this section, we concentrate on solving the above listed problems for Alveo U280. We provide some insights to solve the issues for Stratix 10 MX in Section~\ref{sec:insight}.

\subsection{Proposed Solution}

\begin{figure}[t]
\centering
\includegraphics[width=0.99\linewidth]{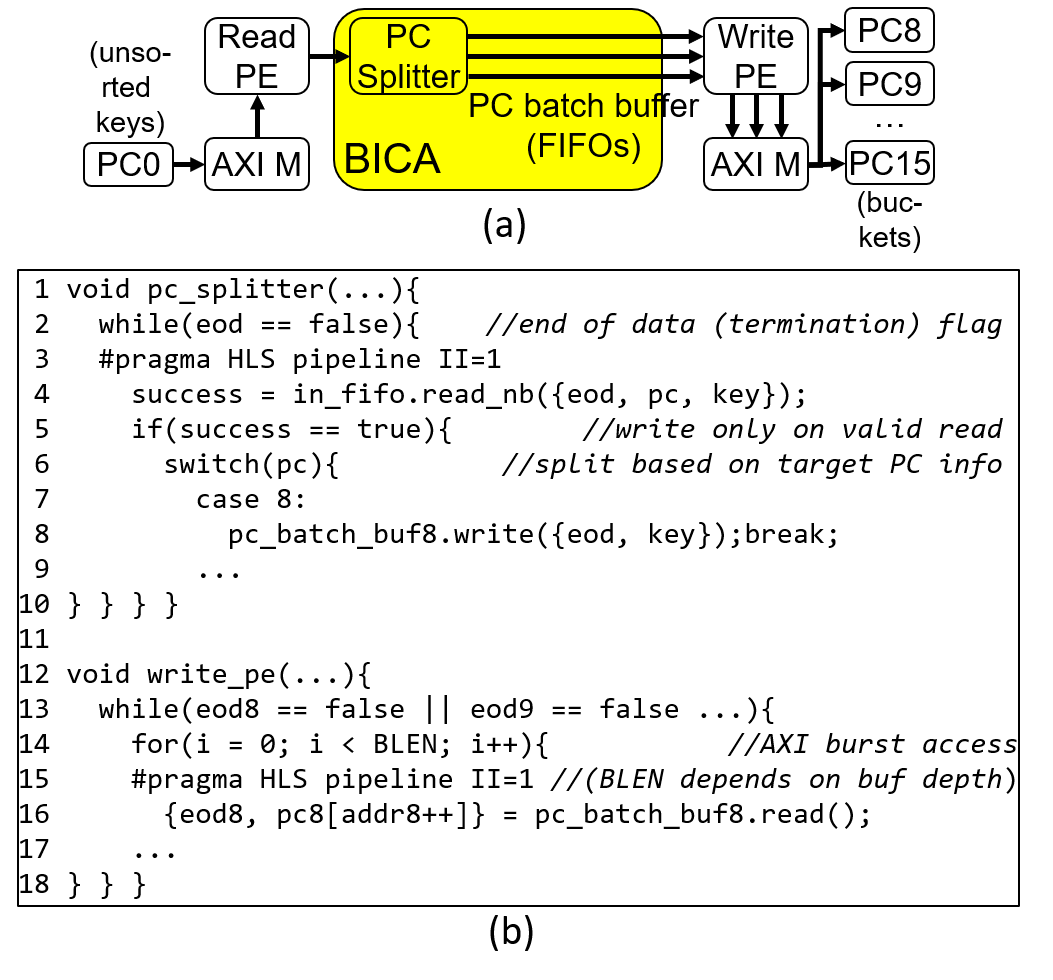}
\caption{Batched inter-channel arbitrator to 8~PCs (a) architecture (b) HLS code}
\label{fig:bica}
\end{figure}

\subsubsection{Solution 1: Batched Inter-Channel Arbitrator (BICA)}

Since Vivado HLS does not automate burst access to different PCs, we propose BICA, which batches the access to each PC. We instantiate multiple FIFOs to be used as batch buffers for each PC (Fig.~\ref{fig:bica}(a)). Based on data's target PC information, a splitter is used to send the data to the corresponding batch buffer (lines~1--10 in Fig.~\ref{fig:bica}(b)). To infer burst access, we make the innermost loop of write PE to read from a FIFO and write to a PC for a fixed burst length (lines~14--16 in Fig.~\ref{fig:bica}(b)). The performance increases with longer burst length (at the cost of larger batch buffer size) as will be shown in Section~\ref{sec:opt_exp}.

The PC splitter logic of BICA has an initiation interval (II) of 1 on Alveo U280. 
The time $t_{BUR}$ taken to complete one burst transaction of length $BLEN$ to HBM PC can be modeled as \cite{Choi2017+,Park2004}:
\begin{equation}
t_{BUR}=BLEN*DW/BW_{max} + LAT
\label{eq:bica_mem_model}
\end{equation}
where $DW$ is the data bitwidth (512b), $BW_{max}$ is the maximum effective bandwidth (sequential access bandwidth) of one PC, and $LAT$ is the memory latency. $BW_{max}$ and $LAT$ are obtained from the microbenchmarks in Section~\ref{sec:evaluation}.
The effective bandwidth of a PC after applying BICA ($BW_{PC}$) is estimated as $BW_{PC}=BLEN*DW/t_{BUR}$ after substituting $t_{BUR}$ with Eq.~\ref{eq:bica_mem_model}. The overall effective bandwidth $BW_{eff}$ ($=PC_{num}*BW_{PC}$) cannot exceed the many-to-many unicast bandwidth ($BW_{mc}$) obtained in Section~\ref{sec:many2many}. $BW_{eff}$ is modeled as:
\begin{equation}
BW_{eff}=min(PC_{num}*BLEN*DW/t_{BUR},BW_{mc})
\label{eq:bica_effbw_model}
\end{equation}

\begin{figure}[t]
\centering
\includegraphics[width=0.99\linewidth]{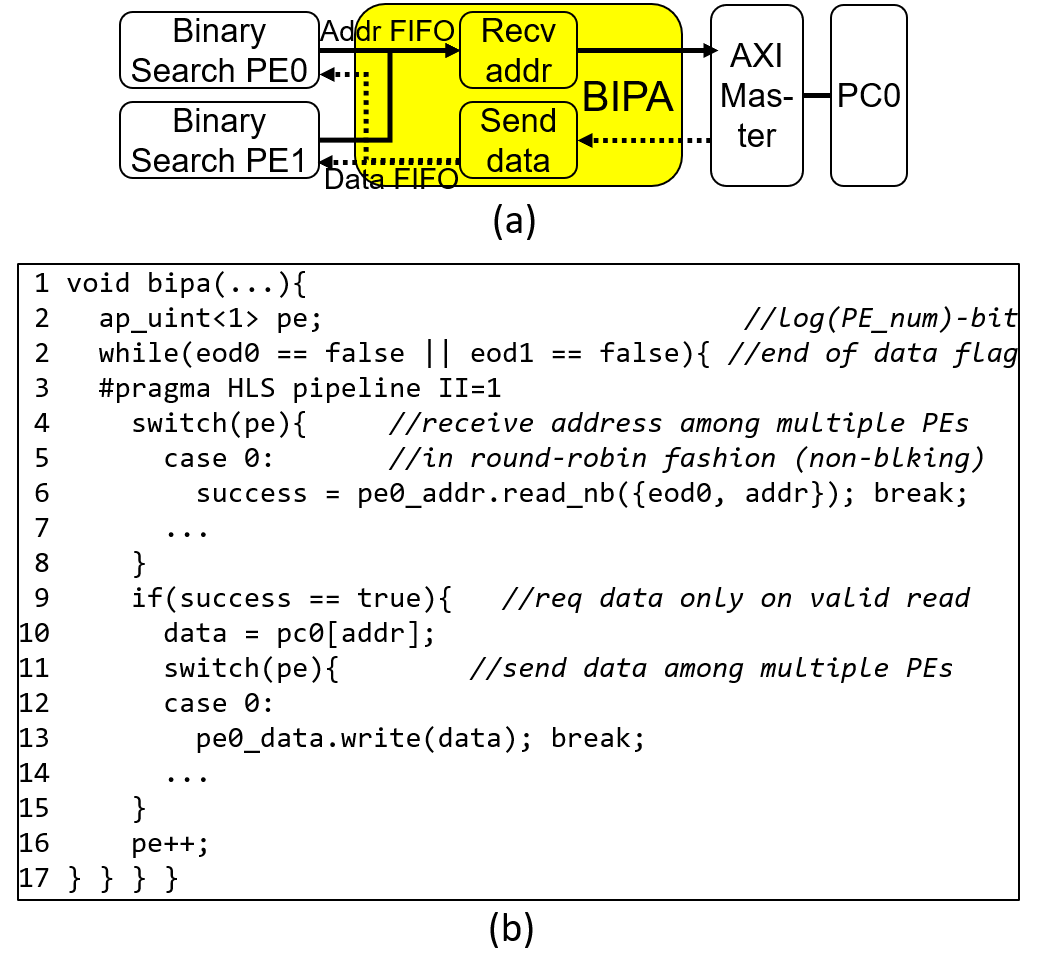}
\caption{Batched inter-PE arbitrator for two PEs (a) architecture (b) HLS code}
\label{fig:bipa}
\end{figure}

\subsubsection{Solution 2: Batched Inter-PE Arbitrator (BIPA)}

In order to allow multiple PEs to share access to a PC in current HBM HLS environment, we propose BIPA, which arbitrates and batches the memory requests from multiple PEs. The architecture of BIPA is shown in Fig.~\ref{fig:bipa}(a). BIPA receives the address of memory requests in a non-blocking, round-robin fashion from the connected PEs (lines~4--8 of Fig.~\ref{fig:bipa}(b)). Then the memory access request is serialized and sent to HBM PC. After receiving the read data from HBM PC, the data is sent back to the PE that has requested it (lines~9--15 of Fig.~\ref{fig:bipa}(b)). For write request arbitration, BIPA can be further simplified by removing the data send logic.

HLS synthesis shows that we can achieve II of 1 for BIPA. From Eq.~\ref{eq:bica_mem_model}, we assume all memory requests have data bitwidth $DW$ and a burst length ($BLEN$) of 1. For $BW_{max}$, we use the strided access bandwidth $BW_{str}$ obtained in Section~\ref{sec:stridebw}. The bandwidth is divided by the number of PEs ($PE_{num}$) accessing BIPA due to the time sharing among PEs ($BW_{max}=BW_{str}/PE_{num}$). The latency $LAT$ is affected by the HMB memory system $LAT_{HBM}$ and PE overhead $LAT_{PE}$ as mentioned in Section~\ref{sec:latency}. It should also include the address and data arbitration latency that increases with the number of PEs. The effective bandwidth using BIPA is modeled as:
\begin{equation}
BW_{eff}=PC_{num}/[1/BW_{str}+(LAT_{HBM} + LAT_{PE} + 2*PE_{num})/DW]
\label{eq:bipa_effbw_model}
\end{equation}

\subsection{Experimental Results}
\label{sec:opt_exp}

\subsubsection{BICA}

Table~\ref{tab:resource_bicabipa} shows the resource consumption of BICA that accesses 8~PCs with 512b data and has FIFO depth of 64.

\begin{table}[ht]
\caption{FPGA resource consumption of BICA and BIPA}
\label{tab:resource_bicabipa}
\centering
\begin{tabular}{|c|c c c c|}
\hline
&LUT & FF & DSP & BRAM \\
\hline
BICA (to 8 PCs) & 5.3K & 11K & 0 & 60 \\
BIPA (from 8 PEs) & 1.4K & 1.4K & 0 & 0 \\
\hline
\end{tabular}
\end{table}

We test the performance improvement from using BICA with bucket sort (Section~\ref{sec:bsort}) and radix sort applications (Section~\ref{sec:rsort}). The result is shown in Table~\ref{tab:bica}.
The effective bandwidth increases with longer burst length. For bucket sort, it is saturated by the many-to-many unicast bandwidth ($BW_{mc}$=96~GB/s, from Table~\ref{tab:crossbar}) at $BLEN$=64 (=4KB). The effective bandwidth of radix sort is slightly lower than that of bucket sort due to the low kernel frequency. The averaged effective bandwidth improvement of the two applications is 2.4X. Note that the BRAM usage does not change with increasing burst length because the minimum depth of Alveo U280 BRAMs is 512. That is, the BRAMs in PC batch buffers are under-utilized.

\begin{table}[ht]
\caption{Effective bandwidth and FPGA resource consumption of bucket sort and radix sort on Alveo U280}
\label{tab:bica}
\centering
\begin{tabular}{|c|c c|c|c c|}
\hline
\multirow{2}{*}{App} & \multirow{2}{*}{Arch}&B & FPGA Resource&KClk&EffBW\\
& &LEN& LUT/FF/DSP/BRAM &(MHz)&(GB/s)\\
\hline
& \multicolumn{2}{c|}{Baseline} & 42K / 108K / 0 / 124 & 300 & 36\\
Buc & & 16 & 102K / 192K / 0 / 604 & 242 & 58\\
sort & BICA & 32 & 103K / 192K / 0 / 604 & 232 & 88\\
 & & 64 & 102K / 192K / 0 / 604 & 220 & 98\\
\hline
& \multicolumn{2}{c|}{Baseline} &  111K / 223K / 0 / 124 & 217 & 35\\
Rad & & 16  &  170K / 286K / 0 / 604 & 191 & 42\\
sort & BICA & 32 &  169K / 278K / 0 / 604 & 200 & 68\\
 & & 64 & 169K / 278K / 0 / 604 & 180 & 75\\
\hline
\end{tabular}
\end{table}

\subsubsection{BIPA}

In Table~\ref{tab:resource_bicabipa}, we report the resource consumption of BIPA that allows 8~PEs with 512b data to share access to a single PC. We test the performance improvement from using BIPA with binary search (Section~\ref{sec:bsearch}) and depth-first search (Section~\ref{sec:dfs}). The result is shown in Table~\ref{tab:bipa}. As more PEs are added, the resource consumption increases as well.
The effective bandwidth does not improve linearly with more PEs because of the arbitration overhead (Eq.~\ref{eq:bipa_effbw_model}). Compared to the baseline implementation of 1 PE per PC, the averaged effective bandwidth improvement of the two applications is 3.8X.

\begin{table}[ht]
\caption{Effective bandwidth and FPGA resource consumption of binary search and depth-first search on Alveo U280}
\label{tab:bipa}
\centering
\begin{tabular}{|c|c c|c|c c|}
\hline
\multirow{2}{*}{App} & \multirow{2}{*}{Arch}&PE & FPGA Resource&KClk&EffBW\\
& &\#& LUT/FF/DSP/BRAM &(MHz)&(GB/s)\\
\hline
& \multicolumn{2}{c|}{Baseline} & 35K / 54K / 0 / 46 & 300 & 7.7\\
\multirow{2}{*}{Bin} & & 2 & 66K / 98K / 0 / 50 & 300 & 11\\
\multirow{2}{*}{srch} & \multirow{2}{*}{BIPA} & 4 & 113K / 162K / 0 / 57 & 274 & 17\\
 & & 8 & 204K / 280K / 0 / 57 & 237 & 23 \\
 & & 16 & 379K / 518K /112/ 57 & 135 & 34 \\
\hline
& \multicolumn{2}{c|}{Baseline} & 35K / 70K / 0 / 88 & 300 & 7.4\\
\multirow{2}{*}{DFS}& & 2 & 63K / 123K / 0 / 106 & 300 & 11\\
& BIPA & 4 & 102K / 190K / 0 / 141 & 284 & 17\\
& & 8 & 182K / 308K / 0 / 197 & 199 & 23\\
\hline
\end{tabular}
\end{table}

\section{Insights for HBM HLS Improvement}
\label{sec:insight}

In this section, we provide a list of insights for HLS tool vendors or researchers to further improve the existing FPGA HBM design flow.

\noindent \textit{\textbf{Insight 1}: Customizable Crossbar Template}

The FPGA resource need to enable many-to-many unicast architecture was kept small for Alveo U280 because of the pre-defined AXI crossbar (Section~\ref{sec:many2many}). But the effective bandwidth was reduced beyond 4$\times$4 AXI masters/slaves. By increasing the complexity of the unit crossbar to 8$\times$8 or by employing an additional stage of a crossbar, we could obtain larger effective bandwidth at the cost of increased resource. On Stratix 10 MX, area rapidly increases when accessing multiple PCs, and it was only possible to fully access up to 8 output PCs (16$\times$16 crossbar failed routing). In both platforms, it would significantly improve the performance and reduce the design time if HLS vendors provide highly optimized and customizable templates of the HBM crossbar.

\smallskip

\noindent \textit{\textbf{Insight 2}: Virtual Channel for HLS}

In BICA (Section~\ref{sec:opt}), many different (and under-utilized) FIFOs are needed to enable burst access into different HBM2 PCs. But such architecture increases the number of FIFO control logic and complicates the PnR process. One possible solution is to allow a single physical FIFO to be shared among multiple virtual channels \cite{Dally1987} in HLS. A new HLS syntax would be needed to allow FIFO access with a virtual channel tag. This is likely to make it easier to share the FIFO control logic and thus reduce the FPGA resource consumption.

\smallskip

\noindent \textit{\textbf{Insight 3}: Low-Latency Memory Pipeline}

Since FPGA HBM platforms are likely to be used for various memory-bounded applications, it is important that HBM HLS tools also support memory latency-bounded applications. We have demonstrated with pointer-chasing microbenchmark (Section~\ref{sec:latency}) and the binary search application that the deep memory pipeline of current HLS tools degrade the performance of latency-bounded applications. It would help programmers if HLS tools provide an option of using less pipelined memory system (possibly at the cost of slightly degraded throughput).

\smallskip

\section{Conclusion}

Evaluation result of microbenchmark and application shows that we can achieve effective bandwidth of 310\textasciitilde390GB/s in recent HBM2 FPGA board. This allows various memory-bound applications to achieve high throughput. But due to the overhead and the architectural limitation of the logic generated by HLS tools, it is sometimes difficult to achieve high performance. The novel HLS-based optimization techniques presented in this paper improve the effective bandwidth when a PE accesses multiple PCs or when multiple PEs share access to a PC. Our research also provides insights to further improve the HBM HLS design flow.

\section{Acknowledgments}

This research is in part supported by Intel and NSF Joint Research Center on Computer Assisted Programming for Heterogeneous Architectures (CAPA) (CCF-1723773), NSF Grant on RTML: Large: Acceleration to Graph-Based Machine Learning (CCF-1937599), Xilinx Adaptive Compute Cluster (XACC) Program, and Google Faculty Award. We thank Michael Adler, Aravind Dasu, John Freeman, Lakshmi Gopalakrishnan, Mohamed Issa, Audrey Kertesz, Eriko Nurvitadhi, Manan Patel, Hongbo Rong, Oliver Tan at Intel, Thomas Bollaert, Matthew Certosimo, and David Peascoe at Xilinx for helpful discussions and suggestions.

\bibliographystyle{ACM-Reference-Format}
\bibliography{sample-base}


\end{document}